\documentclass[aps,prc,twocolumn,superscriptaddress,showpacs]{revtex4}

\usepackage{graphicx}
\usepackage{longtable}
\usepackage{dcolumn}
\usepackage{bm}
\usepackage{ulem}
\usepackage{color}
\usepackage{epsfig}

\begin{document}

\title{
Resonance strengths in the $^{14}$N(p,$\gamma$)$^{15}$O astrophysical key reaction measured with activation
}
\author{Gy. Gy\"urky}%
\email{gyurky@atomki.mta.hu}
\affiliation{Institute for Nuclear Research (Atomki), H-4001 Debrecen, Hungary}
\author{Z. Hal\'asz}%
\affiliation{Institute for Nuclear Research (Atomki), H-4001 Debrecen, Hungary}
\author{G.G.~Kiss}%
\affiliation{Institute for Nuclear Research (Atomki), H-4001 Debrecen, Hungary}
\author{T. Sz\"ucs}
\affiliation{Institute for Nuclear Research (Atomki), H-4001 Debrecen, Hungary}
\author{A. Cs\'ik}%
\affiliation{Institute for Nuclear Research (Atomki), H-4001 Debrecen, Hungary}
\author{Zs.~T\"or\"ok}%
\affiliation{Institute for Nuclear Research (Atomki), H-4001 Debrecen, Hungary}
\author{R. Husz\'ank}%
\affiliation{Institute for Nuclear Research (Atomki), H-4001 Debrecen, Hungary}
\author{M.G. Kohan}%
\affiliation{Department of Engineering Sciences and Mathematics, Lule\aa~University of Technology, 97187 Lule\aa, Sweden}
\author{L. Wagner}%
\altaffiliation{Present address: National Superconducting Cyclotron Laboratory, Michigan State University, East Lansing, USA}
\affiliation{Helmholtz-Zentrum Dresden-Rossendorf, Germany}
\affiliation{Technische Universit\"at Dresden, Germany}
\author{Zs. F\"ul\"op}%
\affiliation{Institute for Nuclear Research (Atomki), H-4001 Debrecen, Hungary}

\date{\today}

\begin{abstract}
\begin{description}

\item[Background]

The $^{14}$N(p,$\gamma$)$^{15}$O reaction plays a vital role in various astrophysical scenarios. Its reaction rate must be accurately known in the present era of high precision astrophysics. The cross section of the reaction is often measured relative to a low energy resonance, the strength of which must therefore be determined precisely.

\item[Purpose]

The activation method, based on the measurement of $^{15}$O decay, has not been used in modern measurements of the $^{14}$N(p,$\gamma$)$^{15}$O reaction. The aim of the present work is to provide strength data for two resonances in the $^{14}$N(p,$\gamma$)$^{15}$O reaction using the activation method. The obtained values are largely independent from previous data measured by in-beam gamma-spectroscopy and are free from some of their systematic uncertainties.  

\item[Method]

Solid state TiN targets were irradiated with a proton beam provided by the Tandetron accelerator of Atomki using a cyclic activation. The decay of the produced $^{15}$O isotopes was measured by detecting the 511\,keV positron annihilation $\gamma$-rays.

\item[Results]

The strength of the E$_p$\,=\,278\,keV resonance was measured to be $\omega\gamma_{278}$\,=\,(13.4\,$\pm$\,0.8)\,meV while for the E$_p$\,=\,1058\,keV resonance $\omega\gamma_{1058}$\,=\,(442\,$\pm$\,27)\,meV. 

\item[Conclusions]

The obtained E$_p$\,=\,278\,keV resonance strength is in fair agreement with the values recommended by two recent works. On the other hand, the E$_p$\,=\,1058\,keV resonance strength is about 20\,\% higher than the previous value. The discrepancy may be caused in part by a previously neglected finite target thickness correction. As only the low energy resonance is used as a normalization point for cross section measurements, the calculated astrophysical reaction rate of the $^{14}$N(p,$\gamma$)$^{15}$O reaction and therefore the astrophysical consequences are not changed by the present results.

\end{description}
\end{abstract}


\maketitle

\section{Introduction}
\label{sec:intro}

Catalytic cycles of hydrogen burning represent an alternative way to the pp-chains for converting four protons into one alpha particle in stellar interiors and for providing thus the energy source of stars. The simplest cycle is the first CNO or Bethe-Weizs\"acker cycle \cite{Wiescher2018} where carbon, nitrogen and oxygen isotopes are involved. The CNO cycle is the dominant energy source of main sequence stars more massive than about 1.3 solar masses but it also plays an important role in various astrophysical scenarios including quiescent and explosive burning processes \cite[e.g.]{DEGLINNOCENTI200413,Fields_2018}.

In the 21$^{\rm th}$ century, astronomical observations as well as astrophysical models are becoming more and more precise. The insufficient knowledge of nuclear reaction rates often represent the largest uncertainty of stellar models. Increasing the precision of experimental nuclear cross sections is thus needed in order to provide accurate reaction rates for the models. For solar models, for example, a precision well below 5\,\% is required for the $^{14}$N(p,$\gamma$)$^{15}$O reaction discussed in the present work \cite{Haxton_2008}.

The slowest reaction of the CNO cycle is the radiative proton capture of $^{14}$N and therefore the rate of this $^{14}$N(p,$\gamma$)$^{15}$O reaction determines the rate of the cycle and hence its efficiency and its contribution to the stellar energy generation. Realizing its importance, many experiments have been devoted to the measurement of its cross section (the full list of references can be found in \cite{RevModPhys.83.195} and the three latest sets of results are published in \cite{PhysRevC.94.025803,PhysRevC.93.055806,PhysRevC.97.015801}). 

Depending on the astrophysical site, the relevant temperature where the CNO cycle is active and important is between about 15 and 200\,MK. This translates into astrophysically relevant center-of-mass energy ranges (the Gamow-window) of the $^{14}$N(p,$\gamma$)$^{15}$O reaction between 20 and 200 keV. Measured cross sections are available only down to 70\,keV. Consequently, for lower temperature environments (like for example our Sun with its 15.7\,MK core temperature) theoretical cross sections or extrapolation of the available data are necessary.

At low energies the $^{14}$N(p,$\gamma$)$^{15}$O reaction proceeds mostly through the direct capture mechanism with contribution from wide resonances. The total cross section is dominated by the capture to the $E_x$\,=\,6.79\,MeV excited states in $^{15}$O but the capture to the ground state and to the $E_x$\,=\,6.17\,MeV excited state also contributes significantly. At higher energies, on the other hand, where cross section data are available, transitions to other states as well as narrow and wide resonances play important roles. 

\subsection{The importance of the $E_p$\,=\,278\,keV resonance}
\label{subsec:278kev_res}

The extrapolation of the cross section to astrophysical energies is typically carried out using the R-matrix approach. For a reliable R-matrix extrapolation, high precision experimental data in a wide energy range is needed for all the relevant transitions and for the resonances as well as for the direct capture. 

At $E$\,=\,259\,keV center-of-mass energy the $^{14}$N(p,$\gamma$)$^{15}$O reaction exhibits a strong narrow resonance. Its energy is too high for any direct astrophysical relevance, however, it plays an important role in the experiments targeting the $^{14}$N(p,$\gamma$)$^{15}$O cross section measurement. In direct kinematics this resonance is observed at $E_p$\,=\,278\,keV proton energy and often serves as a normalization point for the measured non-resonant cross section data, i.e. the cross section is measured relative to the strength of this resonance. Therefore, the precision of this resonance strength directly influences the precision of the cross section data at low energies. 

The available measured  $E_p$\,=\,278\,keV resonance strength values are summarized in Table\,\ref{tab:278strength_literature}. Based on several measurements, the recommended value of the strength has an uncertainty of 4.6\,\%, as given by the latest compilation of solar fusion reactions \cite{RevModPhys.83.195}. In the paper of Daigle \textit{et al.} \cite{PhysRevC.94.025803} published after the above cited compilation, a new result was presented and the literature data were also critically re-analyzed. Their recommended value is in agreement with that of \cite{RevModPhys.83.195} but its uncertainty is reduced to 2.4\,\%. This uncertainty seems surprisingly low considering on one hand the stopping power uncertainty which is common to almost all the experiments and on the other hand the difficulty in characterizing the implanted targets used by Daigle \textit{et al.} \cite{PhysRevC.94.025803}.

In the present work the $E_p$\,=\,278\,keV resonance strength is measured using an independent method, the activation technique \cite{Gyurky2019}. As an additional result, the strength of the $E_p$\,=\,1058\,keV resonance in $^{14}$N(p,$\gamma$)$^{15}$O reaction is also measured. This resonance plays no role in astrophysics, but it can also be used as a reference point for the $^{14}$N(p,$\gamma$)$^{15}$O non-resonant cross section measurements and an R-matrix extrapolation of experimental data also requires the knowledge of the parameters of this resonance. Although this resonance is stronger than the $E_p$\,=\,278\,keV one, its strength has been measured in fewer experiments (see Table\,\ref{tab:1058strength_literature}) and the precision of the strength recommended by Marta \textit{et al.} is not better than about 5\,\%.

\begin{table*}
\caption{\label{tab:278strength_literature} Available $E_p$\,=\,278\,keV resonance strength data in the literature}
\begin{ruledtabular}
\begin{tabular}{lllr@{\,}c@{\,}l}
Reference & Year  & Method & \multicolumn{3}{c}{Resonance strength}\\
          &       &        & \multicolumn{3}{c}{$\omega\gamma_{278}$/meV}\\
\hline
E.J. Woodbury  {\it et al.} \cite{woodbury1949}	 &	1949	 &	activation	 &	\multicolumn{3}{c}{10}					\\
D.B. Duncan, J.E. Perry	\cite{PhysRev.82.809} &	1951	 &	activation	 &	\multicolumn{3}{c}{20}					\\
S. Bashkin {\it et al.} \cite{PhysRev.99.107}	 &	1955	 &	prompt gamma	 &	13	 &	 $\pm$	 &	3	\\
D.F. Hebbard {\it et al.} \cite{HEBBARD1963666}	 &	1963	 &	prompt gamma	 &	14	 &	 $\pm$	 &	2	\\
H.W. Becker {\it et al.} \cite{Becker1982}	 &	1982	 &	prompt gamma	 &	14	 &	 $\pm$	 &	1	\\
R. C. Runkle {\it et al.} \cite{PhysRevLett.94.082503}	 &	2005	 &	prompt gamma	 &	13.5	 &	 $\pm$	 &	1.2	\\
G. Imbriani  {\it et al.} \cite{Imbriani2005}	 &	2005	 &	prompt gamma	 &	12.9	 &	 $\pm$	 &	0.9	\\
D. Bemmerer  {\it et al.} \cite{BEMMERER2006297}	 &	2006	 &	prompt gamma	 &	12.8	 &	 $\pm$	 &	0.6	\\
S. Daigle {\it et al.} \cite{PhysRevC.94.025803}	 &	2016	 &	prompt gamma	 &	12.6	 &	 $\pm$	 &	0.6	\\
\hline
recommended by E.G. Adelberger {\it et al.} \cite{RevModPhys.83.195}	 &	2011	 &		 &	13.1	 &	 $\pm$	 &	0.6	\\
recommended by S. Daigle {\it et al.} \cite{PhysRevC.94.025803}	 &	2016	 &		 &	12.6	 &	 $\pm$	 &	0.3	\\
\end{tabular}
\end{ruledtabular}
\end{table*}

\begin{table*}
\caption{\label{tab:1058strength_literature} Available $E_p$\,=\,1058\,keV resonance strength data in the literature}
\begin{ruledtabular}
\begin{tabular}{lllr@{\,}c@{\,}l}
Reference & Year  & Method & \multicolumn{3}{c}{Resonance strength}\\
          &       &        & \multicolumn{3}{c}{$\omega\gamma_{1058}$/meV}\\
\hline
D.B. Duncan, J.E. Perry	\cite{PhysRev.82.809} &	1951	 &	activation	 &	\multicolumn{3}{c}{630}					\\
D.F. Hebbard {\it et al.} \cite{HEBBARD1963666}	 &	1963	 &	prompt gamma	 &	\multicolumn{3}{c}{394}					\\
U. Schr\"oder   {\it et al.} \cite{SCHRODER1987240} &	1987	 &	prompt gamma	 &	310	 &	 $\pm$	 &	40	\\
M. Marta {\it et al.} \cite{PhysRevC.81.055807} &	2010	 &	prompt gamma	 &	364	 &	 $\pm$	 &	21	\\
\hline
recommended by M. Marta  {\it et al.} \cite{PhysRevC.81.055807}	 &	2010	 &		 &	353	 &	 $\pm$	 &	18	\\
\end{tabular}
\end{ruledtabular}
\end{table*}

\subsection{The activation method for the study of the $^{14}$N(p,$\gamma$)$^{15}$O reaction}
\label{subsec:actmethod}

The proton capture of $^{14}$N at the $E_p$\,=\,278\,keV and $E_p$\,=\,1058\,keV resonances leads to the formation of $^{15}$O in excited states of $E_x$\,=\,7556\,keV and $E_x$\,=\,8284\,keV, respectively. These excited states decay to the $^{15}$O ground state by the emission of prompt $\gamma$-radiation through several possible cascades. The detection of this $\gamma$-radiation has been used in almost all the experiments to determine the resonance strengths (see the entries in tables \ref{tab:278strength_literature} and \ref{tab:1058strength_literature} labeled as 'prompt gamma'). For a precise resonance strength measurement the $\gamma$-detection efficiency (up to 8\,MeV $\gamma$-energy), the angular distribution of the various transitions as well as the branching ratio of these transitions (including the weak ones) must be known precisely. All these factors introduce systematic uncertainties in the resonance strength determination.

Since the reaction product of the $^{14}$N(p,$\gamma$)$^{15}$O reaction is radioactive, the resonance strength can also be measured by activation. $^{15}$O decays by positron emission to $^{15}$N with a half-life of 122.24\,$\pm$\,0.16\,s \cite{AJZENBERGSELOVE19911}. The decay is not followed by the emission of $\gamma$-radiation, however, the 511\,keV $\gamma$-ray following the positron annihilation provides a possibility for the reaction strength measurement with activation. This method is free from some uncertainties encumbering the prompt gamma experiments. The decay occurs isotropically, thus no angular distribution needs to be measured. The $\gamma$-detection efficiency must be known only at a single, low energy point (511\,keV) where it is measured more easily than at several MeV's. Since by the activation method the number of produced isotopes is measured, this technique provides directly the total reaction cross section or resonance strength (independent from the decay scheme of the exited levels) and therefore no uncertainty arises from weak transitions.

The activation method was used only by some very first studies of the $^{14}$N(p,$\gamma$)$^{15}$O reaction about 70 years ago (see tables \ref{tab:278strength_literature} and \ref{tab:1058strength_literature}) and these measurements did not lead to precise resonance strengths. In the present work this method is used again to provide precise resonance strength values which are largely independent from the ones measured with prompt gamma detection. The next section provides details of the experimental procedure while the data analysis is presented in Sec.\,\ref{sec:analysis}. The final results, their comparison with available data and conclusions are given in Sec.\,\ref{sec:results}. 

\section{Experimental procedure}
\label{sec:experiment}

\subsection{Target preparation and characterization}
\label{subsec:target}

In many of the past experiments solid state titanium-nitride (TiN) targets proved to be an excellent choice to carry out ion-beam induced reaction studies on nitrogen isotopes \cite[e.g.]{PhysRevC.97.015801,PhysRevC.93.055806,PhysRevC.81.055807,2004PhLB..591...61F}. TiN can be produced at the required thickness and purity and these targets can withstand intense beam bombardment. The Ti:N atomic ratios are typically found to be very close to 1:1 in such targets.

Considering the advantages, solid state TiN targets were used also in the present work. They were prepared by reactive sputtering of TiN onto 0.5\,mm thick Ta backings at the Helmholtz-Zentrum Dresden-Rossendorf, Germany. The nominal thicknesses of the TiN layers were between 100 and 300\,nm, but for the purpose of the resonance strength measurements presented here, only 300\,nm thick targets were used. This corresponds to roughly 1.5$\times$10$^{18}$ N atoms/cm$^2$. 

For the resonance strength determination, the total thickness of the targets does not play a role (as long as the thick target assumption holds, see Sec.\,\ref{sec:analysis}). The stoichiometry, i.e. the Ti:N ratio, on the other hand, is a crucial parameter as discussed in Sec.\,\ref{sec:analysis}. The Ti:N ratio and the amount of impurities were therefore measured with three independent methods: SNMS, RBS and PIXE, as described below. For these measurements, TiN layers sputtered onto Si wafers were used. These samples had been prepared together with the actual  targets on Ta backings in the same sputtering geometry and therefore the layer compositions are the same.

\begin{figure}
\includegraphics[angle=270,width=\columnwidth]{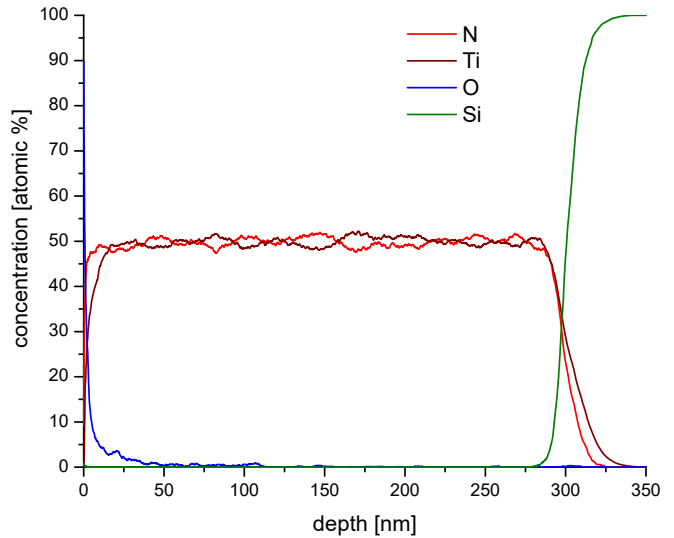}\\
\caption{\label{fig:SNMS} Concentration of the various elements in the target as a function of depth measured with the SNMS technique.}
\end{figure}

Secondary Neutral Mass Spectrometry (SNMS) technique was used to measure the target composition as a function of the depth of the layer. The measurement was done with the INA-X type (SPECS GmbH, Berlin) SNMS facility of Atomki \cite{OECHSNER1993250,vad2009}. Figure\,\ref{fig:SNMS} shows a typical SNMS profile of a 300\,nm target. Besides the small amount of oxygen contamination on the surface, no elements other than nitrogen and titanium  were observed. The Ti:N ratio was found to be uniform along the thickness of the target within the statistical fluctuation of the data and the average ratio is 1.015\,$\pm$\,0.051. The uncertainty includes the statistical component (less than 1\,\%) and a 5\,\% systematic uncertainty.

\begin{figure}
\includegraphics[angle=270,width=\columnwidth]{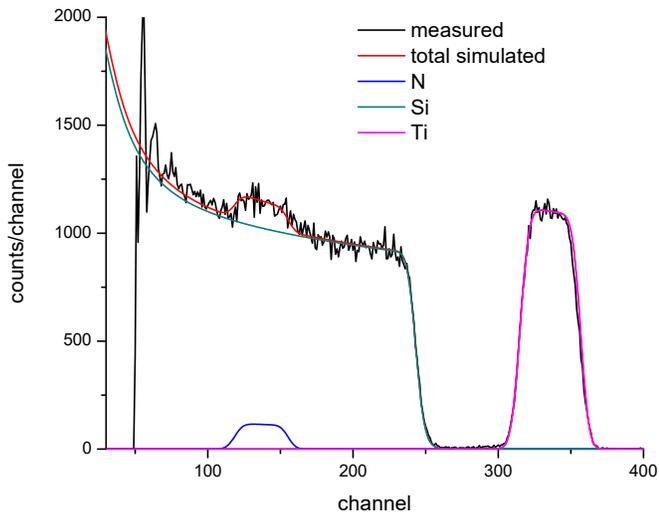}\\
\caption{\label{fig:RBS} A typical measured and simulated RBS spectrum.  }
\end{figure}

Rutherford Backscattering Spectrometry (RBS) was also used to measure the target composition. A 1.6\,MeV $\alpha$-beam provided by the 5\,MV Van de Graaff accelerator of Atomki was focused onto the targets in an Oxford type microbeam setup \cite{Huszank2016}. The scattered $\alpha$-particles were detected by two ion-implanted Si detectors positioned at 135 and 165 degrees with respect to the incoming beam direction. Figure\,\ref{fig:RBS} shows a typical RBS spectrum. Based on the evaluation of the RBS spectra with the SIMNRA code \cite{SIMNRA}, a Ti:N ratio of 0.976\,$\pm$\,0.048 was determined. The uncertainty includes the fit uncertainty and a 3\,\% systematic one characterizing the general accuracy of the used RBS system which was assessed based on the measured thickness reproducibility of several RBS standards.

\begin{figure}
\includegraphics[angle=270,width=\columnwidth]{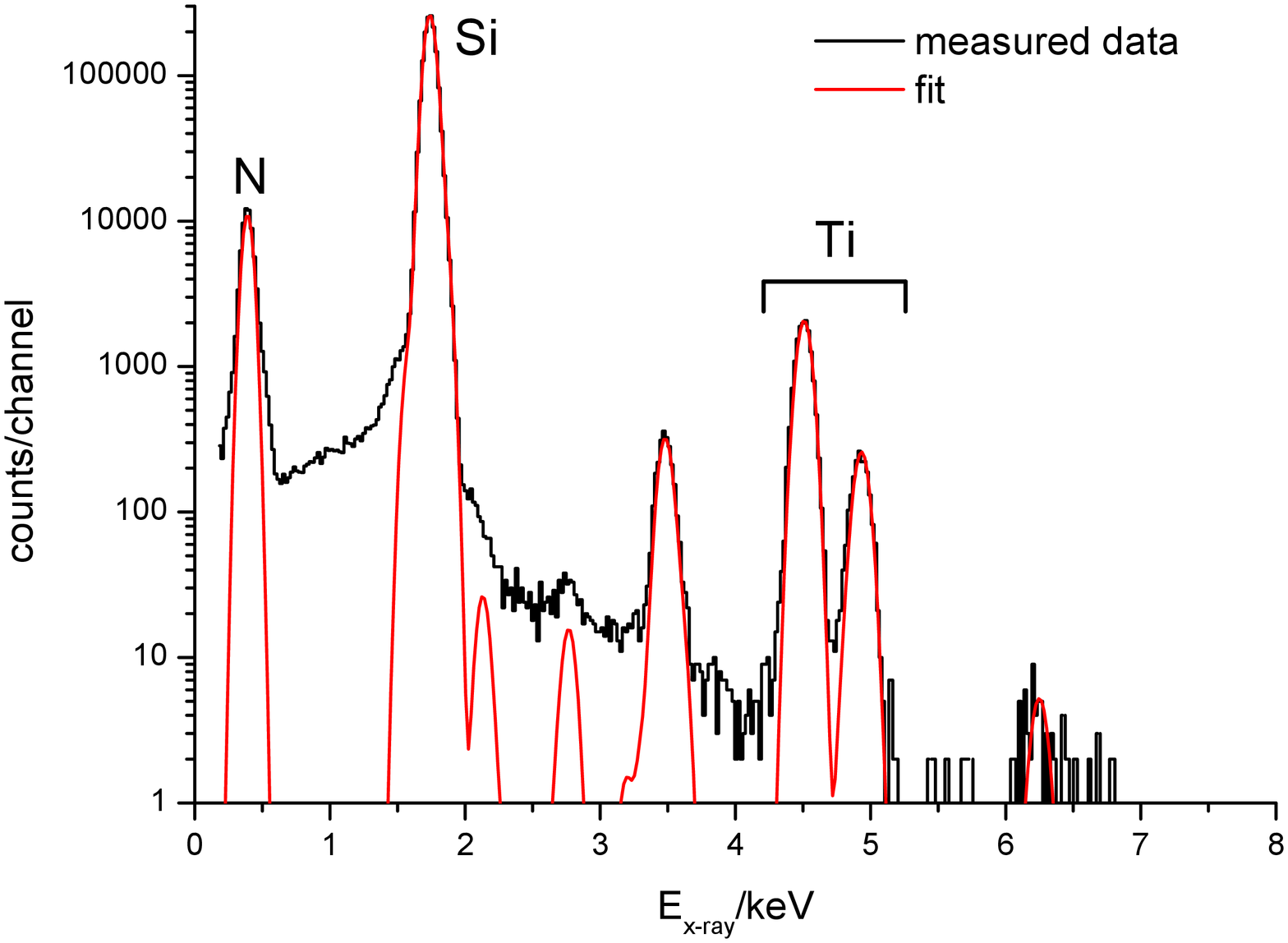}\\
\caption{\label{fig:PIXE} PIXE spectrum of a TiN target. Major elements included in the fit are labeled.}
\end{figure}

Using the same microbeam setup as for the RBS measurement, the targets were also studied with Proton Induced X-ray Emission (PIXE) \cite{Huszank2016}. The targets were bombarded by a 2.0\,MeV proton beam and the induced X-rays were detected by a silicon drift X-ray detector. A typical X-ray spectrum can be seen in Fig.\,\ref{fig:PIXE}. Owing to the thin window of the detector, characteristic X-rays of nitrogen could be detected with good accuracy and a Ti:N ratio of 0.981\,$\pm$\,0.064 was obtained. Here the uncertainty include a 3\,\% systematic component.

Table\,\ref{tab:targets} summarizes the results of the target stoichiometry measurements. All three results are in excellent agreement with the expected 1:1 ratio. Based on the weighted average, 0.991\,$\pm$\,0.031 is adopted as the Ti:N ratio for the resonance strength calculations.

\begin{table}
\caption{\label{tab:targets} Measured Ti:N ratios of the used targets. The adopted ratio is the weighted average of the results of the three methods.}
\begin{ruledtabular}
\begin{tabular}{lrcl}
Method & \multicolumn{3}{c}{Ti:N atomic ratio}\\
\hline
SNMS	 &	1.015	 &	 $\pm$	 &	0.051	\\
RBS	 &	0.976	 &	 $\pm$	 &	0.048	\\
PIXE	&  0.981	 &	 $\pm$	& 0.064	\\
\hline
adopted &	0.991	 &	 $\pm$	 &	0.031	\\
\end{tabular}
\end{ruledtabular}
\end{table}

\subsection{Activations}
\label{subsec:activations}

The proton beams for the excitation of the studied resonances in the $^{14}$N(p,$\gamma$)$^{15}$O reaction were provided by the Tandetron accelerator of Atomki. The energy calibration of the accelerator has been carried out recently \cite{RAJTA2018125} and that was used for setting the energies for the resonance studies. As the resonances are relatively strong, no high beam intensity was necessary which was useful to avoid target deterioration. The typical beam intensity was 5\,$\mu$A on target. 

\begin{figure}
\includegraphics[width=\columnwidth]{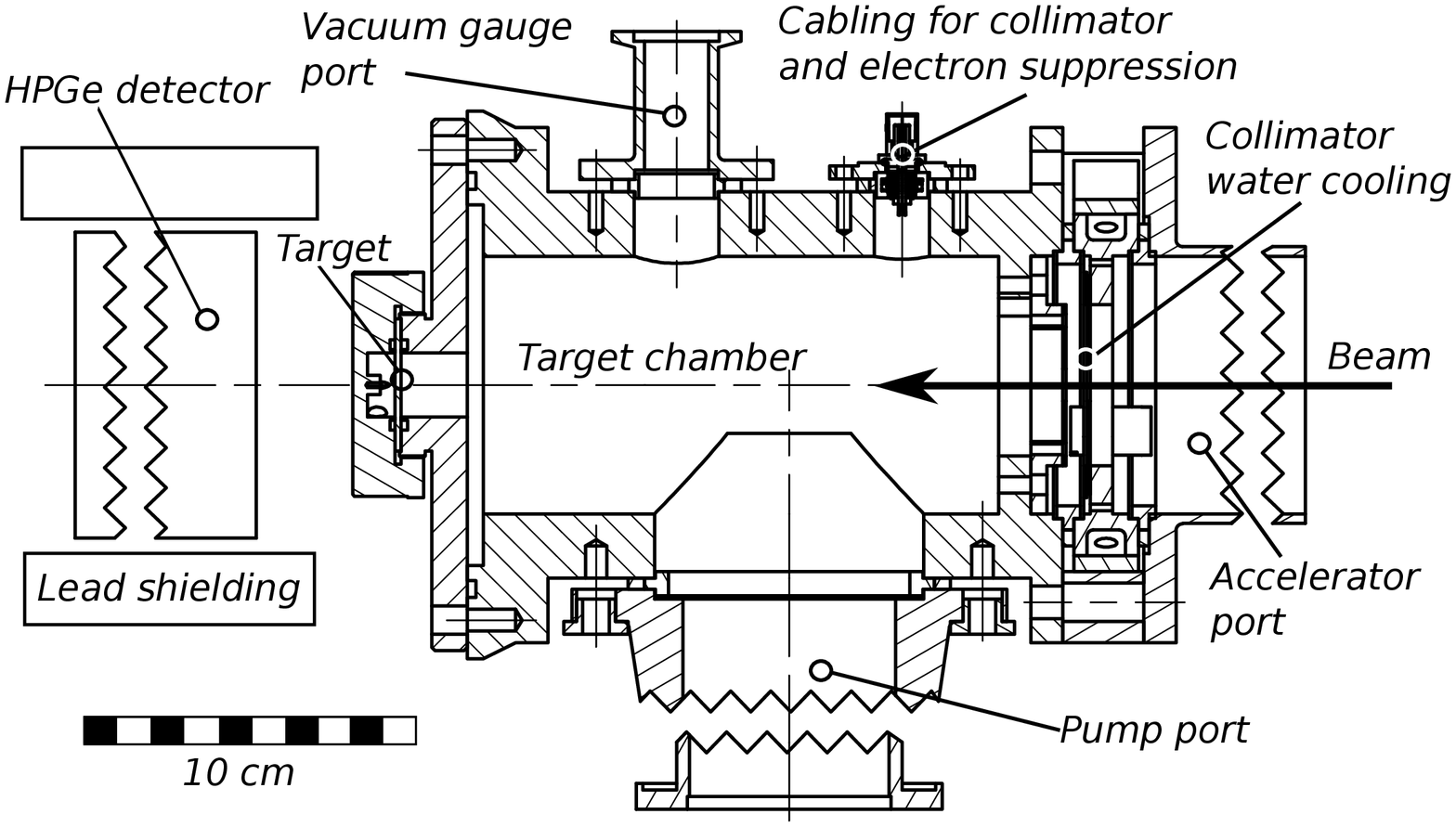}\\
\caption{\label{fig:chamber} Drawing of the target chamber used for the activations}
\end{figure}

The applied beam energies for the study of the 278\,keV and 1058\,keV resonances were $E_p$\,=\,300\,keV and $E_p$\,=\,1070\,keV, respectively. These values correspond to the middle of the yield curve plateau, where the maximum yield can be reached. See the discussion in Sec.\,\ref{sec:analysis}.

The schematic drawing of the target chamber can be seen in Fig.\,\ref{fig:chamber}. The beam enters the chamber through a water cooled collimator of 5\,mm in diameter. Behind the collimator, an electrode biased at -300\,V is placed to suppress secondary electrons emitted from the target or from the collimator. After the collimator the whole chamber serves as a Faraday-cup to measure the charge carried by the beam to the target. The measured charge was used to determine the number of protons impinging on the target.

\subsection{Detection of the annihilation radiation}
\label{subsec:detection}

As the half-life of the $^{15}$O reaction product is rather short (about two minutes),  the induced activity was measured without removing the target from the activation chamber. A 100\,\% relative efficiency HPGe $\gamma$-detector was therefore placed close behind the target. The distance between the target and the detector end-cap was about 1\,cm.  

In order to increase the number of detected decay, the cyclic activation method was applied. The target was irradiated for 5 minutes and then the beam was stopped in the low energy Faraday cup of the accelerator and the decay was measured for 10 or 20 minutes. This cycle was repeated many times (up to 30 cycles in a singe irradiation campaign). 

\begin{figure}
\includegraphics[angle=270,width=\columnwidth]{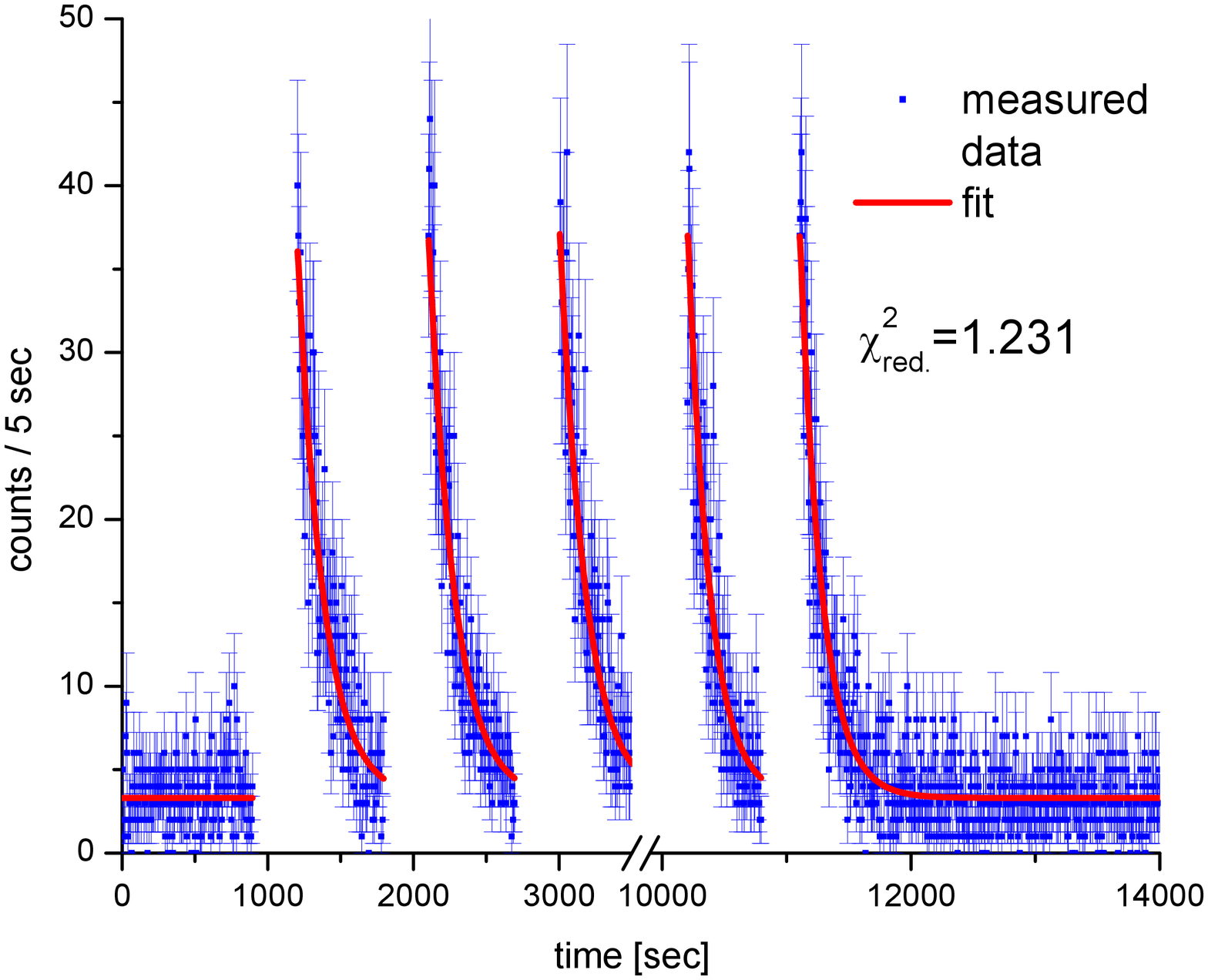}\\
\includegraphics[angle=270,width=\columnwidth]{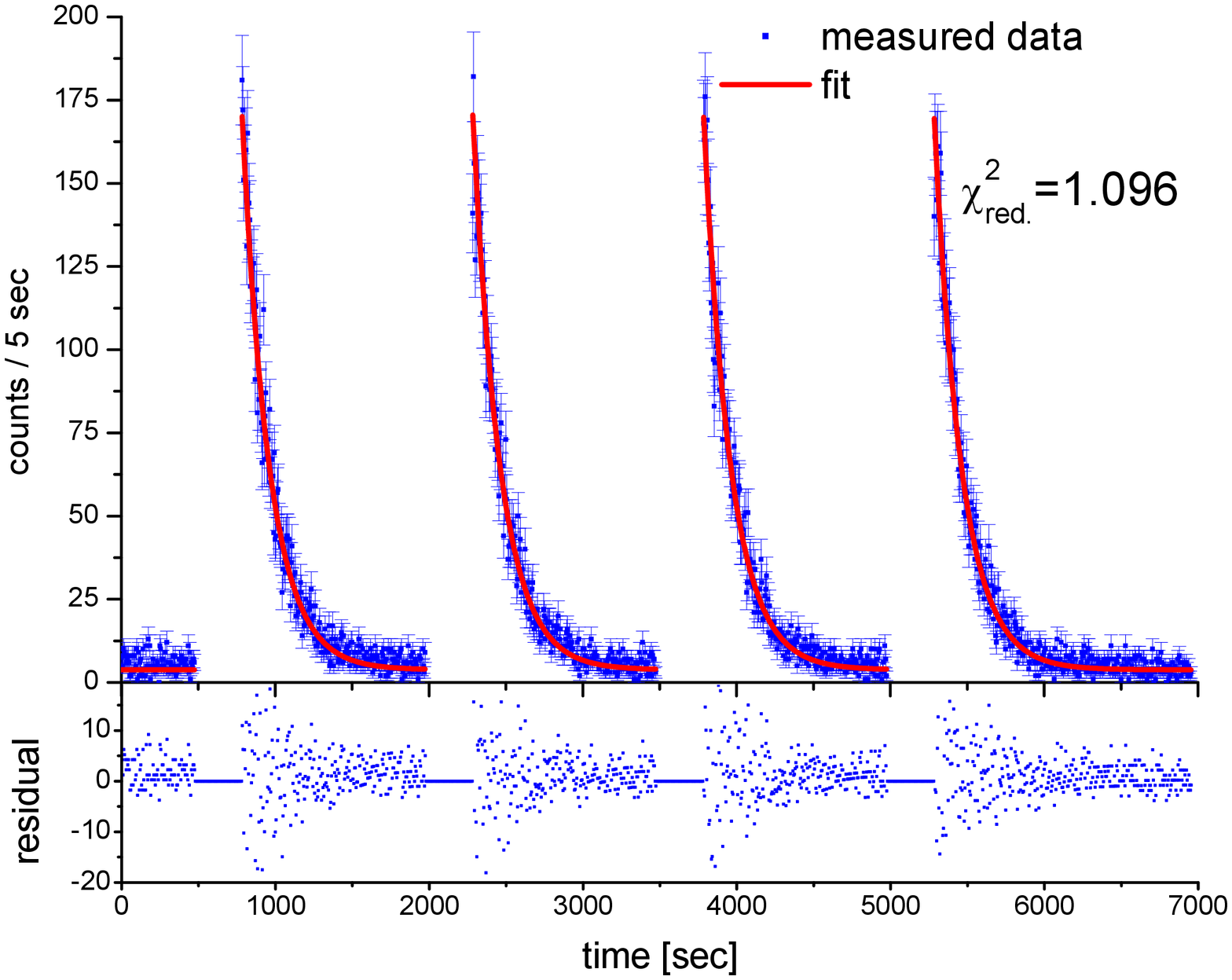}
\caption{\label{fig:decay} Number of events detected by the HPGe detector at the 511\,keV peak region as a function of time using 5 second time bins. The fit to the data including a time-independent laboratory background component and using the known half-life of $^{15}$O is also shown. The upper and lower panels show the $E_p$\,=\,278\,keV and $E_p$\,=\,1058\,keV measurements, respectively. For the latter case the fit residuals are also plotted in order to indicate that the decay of $^{15}$O alone fits well the measured data, no other radioactivity was present in significant amount. This is also confirmed by the reduced $\chi^2$ value of the fit being very close to unity.}
\end{figure}

In order to follow the decay of $^{15}$O, the number of events in the region of 511 keV peak (selected by gating with a single channel analyzer) was recorded in five second time intervals using an ADC in multichannel scaling mode. Figure\,\ref{fig:decay} shows typical examples of the recorded number of counts as a function of time. The upper panel shows a case measured on the $E_p$\,=\,278\,keV resonance with 10 minute counting intervals, while the lower panel represents a measurement on the  $E_p$\,=\,1058\,keV resonance with 20 minute counting intervals. During the 5 minutes irradiation intervals the events in the detector were disregarded as in these periods the counts were dominated by beam induced background.

\subsection{Determination of the detector efficiency}
\label{subsec:efficiency}

For the absolute measurement of the resonance strengths the absolute detection efficiency of the HPGe detector must be known in the counting geometry used. In the present case of a positron decaying isotope, the positron annihilation does not occur in a point-like geometry and hence the precise determination of the detection efficiency is not trivial. 

The positrons leave the decaying $^{15}$O nucleus with typically several hundreds of keV energy (the positron end-point energy is 1732\,keV \cite{AJZENBERGSELOVE19911}). The positrons which travel towards the target backing stop within a few 100\,$\mu$m (thus well inside the backing) and annihilate therefore in a quasi point-like geometry. On the other hand, those positrons which leave towards the other direction, will move into the vacuum chamber and travel freely until they hit the walls of the chamber. Therefore, their annihilation takes place in an extended and not well defined geometry. 

In such a situation the direct efficiency measurement with calibrated radioactive sources is not possible. Instead, an indirect method using the following procedure was applied. As a first step, longer-lived positron emitters were produced in the activation chamber. For this purpose $^{18}$F (t$_{1/2}$\,=\,109.77\,$\pm$\,0.05\,min, produced by the $^{18}$O(p,n)$^{18}$F reaction) and $^{13}$N (t$_{1/2}$\,=\,9.965\,$\pm$\,0.004\,min, produced by the $^{12}$C(p,$\gamma$)$^{13}$N reaction) were chosen. The decay of these sources was measured with the HPGe detector standing next to the chamber (the one which was used for the $^{14}$N(p,$\gamma$)$^{15}$O reaction) for typically 1-2 half-lives. This measurement gives information about the efficiency in the non-trivial extended geometry. Then the sources were removed from the chamber, transferred to another HPGe detector (used in many recent experiments and characterized precisely, see e.g. \cite{PhysRevC.95.035805}) where the decay was followed for several half-lives. At this detector the sources were placed in a position which guaranteed the point-like geometry, i.e. the sources were placed between 0.5\,mm thick Ta sheets which stopped the positrons completely. The absolute efficiency of this HPGe detector in the used geometry was measured with calibrated radioactive sources to a precision of 3\,\%. The measurement with the second detector provided the absolute activity of the sources and -- knowing precisely the half-lives and the elapsed time between the two countings -- the absolute efficiency of the first detector could be obtained. The efficiency obtained with $^{18}$F and $^{13}$N sources were in agreement within the statistical uncertainties of 0.7\,\% and 1.5\,\%, respectively.

In addition to the measurements with $^{18}$F and $^{13}$N, the same procedure was followed also with the actual $^{15}$O isotope produced by the $^{14}$N(p,$\gamma$)$^{15}$O reaction. Here the short half-life resulted in a higher statistical uncertainty of 2.5\,\%, but the obtained efficiency was in agreement with the results from $^{18}$F and $^{13}$N. Based on these measurements the final efficiency is determined with a precision of 4\,\%.

\section{Data analysis}
\label{sec:analysis}

As it can be seen by the red line in Fig.\,\ref{fig:decay}, the 511\,keV count rate can be well fitted with a constant background plus an exponential decay with the $^{15}$O half-life. With the known detection efficiency and the decay parameters of $^{15}$O the only free parameter of the fit is the yield of the reaction (i.e. the number of reactions per incident proton) which can be related to the resonance strength. 

With the thick target assumption (i.e. when the energetic thickness of the target $\Delta E$ is much larger than the natural width $\Gamma$ of the resonance), the resonance strength $\omega\gamma$ can be related to the yield $Y$ measured on the top of the resonance by the following formula:
\begin{equation}
	 \omega\gamma=\frac{2 \epsilon_{\rm eff} Y}{\lambda^2}
\end{equation}
where $\lambda$ is the de Broglie wavelength at the resonance energy in the center of mass system and $\epsilon_{\rm eff}$ is the effective stopping power. If the thick target condition is not met, the maximum  yield $Y_{max}$ measured in the middle of the resonance curve plateau can be related to the ideal thick target yield leading to the following correction factor \cite{RevModPhys.20.236}: 
\begin{equation}
	f \equiv \frac{Y_{max}}{Y} = \frac{2}{\pi} \tan^{-1} \frac{\Delta E}{\Gamma}.
\end{equation}
Based on the target characterizations presented in Sec.\,\ref{subsec:target}, the energetic thickness of the targets used for the present experiments at the two studied resonances was $\Delta E_{278}$\,=\,51.9\,keV and $\Delta E_{1058}$\,=\,24.4\,keV, respectively with about 5\,\% uncertainty. The natural widths of the two resonances (taken from the literature) are $\Gamma_{278}$\,=\,1.12\,$\pm$\,0.03\,keV \cite{doi:10.1063/1.3087064} and $\Gamma_{1058}$\,=\,3.8\,$\pm$\,0.5\,keV  \cite{PhysRevC.81.055807}. These lead to correction factors of $f_{278}$\,=\,0.986\,$\pm$\,0.001 and $f_{1058}$\,=\,0.902\,$\pm$\,0.015. Here the uncertainties take into account the resonance widths, the target thickness and the beam energy uncertainties. The latter one measures how precisely the maximum of the resonance curve is found. It is evident that in the case of the astrophysically more important low energy resonance this correction is very close to unity and its uncertainty is negligible. This is owing to the larger target thickness at this low energy and the small natural width of the resonance.

The effective stopping power $\epsilon_{\rm eff}$ in the case of a target composed of Ti and N can be obtained as
\begin{equation}
	\epsilon_{\rm eff}=\epsilon_{\rm N} + \frac{N_{\rm Ti}}{N_{\rm N}}\epsilon_{\rm Ti} 
\end{equation}
where $\epsilon_{\rm N}$ and $\epsilon_{\rm Ti}$ are the stopping powers of N and Ti, respectively, taken at the resonance energy and ${N_{\rm Ti}}/{N_{\rm N}}$ is the Ti:N atomic ratio as discussed in Sec.\,\ref{subsec:target}.

The stopping power was taken from the 2013 version of the SRIM code \cite{SRIM}. The following values were used: 
\begin{itemize}
	\item $\epsilon_{\rm N}$(278 keV)\,=\,10.72\,eV/(10$^{15}$ atoms/cm$^2$),
	\item $\epsilon_{\rm N}$(1058 keV)\,=\,4.733\,eV/(10$^{15}$ atoms/cm$^2$),
	\item $\epsilon_{\rm Ti}$(278 keV)\,=\,22.81\,eV/(10$^{15}$ atoms/cm$^2$),
	\item $\epsilon_{\rm Ti}$(1058 keV)\,=\,10.80\,eV/(10$^{15}$ atoms/cm$^2$).
\end{itemize}

As for Ti, the stopping power was measured by N. Sakamoto \textit{et al.} \cite{SAKAMOTO2000250} with very good precision of better than 1\,\%. The measured values are in excellent agreement with the SRIM data, never deviating more than 2\,\%, Therefore, an uncertainty of 2\,\% is assigned to $\epsilon_{\rm Ti}$ in the present work. The situation is somewhat worse in the case of nitrogen. The stopping power measured with gaseous N can be different from the solid form (see e.g. \cite{REITER1987287} for the stopping power dependence on the chemical form). Therefore, 4\,\% uncertainty is assigned to $\epsilon_{\rm N}$ as recommended by SRIM. The uncertainty of the effective stopping power was calculated taking into account the uncertainty of the measured Ti:N ratio and considering the $\epsilon_{\rm Ti}$ and $\epsilon_{\rm N}$ values uncorrelated. The isotopic abundance of $^{14}$N in natural nitrogen (99.6337\,\%) was taken into account in the calculation of $\epsilon_{\rm eff}$.

For the determination of the resonance strength, the non-resonant component of the reaction yield must be subtracted from the resonant yield. In the case of the $E_p$\,=\,1058\,keV resonance the yields below and above the resonance -- at $E_p$\,=\,1000\,keV and $E_p$\,=\,1150\,keV -- were measured. Based on these measurements a 2.7\,\% nonresonant contribution to the resonant yield was determined. This non-resonant yield was subtracted from the resonant yield and a conservative relative uncertainty of 20\,\% was assigned to it, leading to a 0.6\,\% uncertainty of the determined resonance strength.

In the case of the $E_p$\,=\,278\,keV resonance the off-resonant reaction yield was below the detection limit. Based on some recent experiments, a cross section of about 1.5$\times10^{-8}$ barns can be expected at this energy \cite{PhysRevC.93.055806,PhysRevC.97.015801}. Such a cross section leads to a calculated non-resonant yield which is 0.3\,\% of the resonant yield. This tiny contribution is subtracted from the yield and - as this value is not based on our own measurement - a 100\,\% relative uncertainty is assigned to it.

Table\,\ref{tab:uncert} lists the uncertainties of the final resonance strength values. As the studied resonances are relatively strong and the cyclic activations were carried out many times, the statistical uncertainty of the $\gamma$-counting is very low compared to the other sources of uncertainty. The quoted total uncertainty is the quadratic sum of the components. Other uncertainties (like for example the uncertainties of the $^{15}$O decay parameters) are well below 1\,\% and are therefore neglected.

\begin{table}
\caption{\label{tab:uncert} Components of the resonance strength uncertainties}
\begin{ruledtabular}
\begin{tabular}{lcc}
Source & 278\,keV & 1058\,keV\\
       & \multicolumn{2}{c}{resonance} \\
\hline
counting statistics &1.0\,\% & 0.7\,\% \\
effective stopping power & \multicolumn{2}{c}{2.8\,\%} \\
HPGe detector efficiency & \multicolumn{2}{c}{4.0\,\%} \\
current integration & \multicolumn{2}{c}{3.0\,\%} \\
finite target thickness correction & 0.1\,\% & 1.7\,\% \\
non-resonant yield subtraction & 0.3\,\% & 0.6\,\% \\
\hline
total uncertainty &  5.8\,\% & 6.0\,\% \\
\end{tabular}
\end{ruledtabular}
\end{table}

\section{Results and conclusions}
\label{sec:results}

The obtained strengths of the two studied resonances are the following:
\begin{itemize}
	\item $\omega\gamma_{278}$\,=\,(13.4\,$\pm$\,0.8)\,meV,
	\item $\omega\gamma_{1058}$\,=\,(442\,$\pm$\,27)\,meV.
\end{itemize}
If we take into account the total uncertainties, the new result for the $E_p$\,=\,278\,keV resonance strength is in good agreement with the adopted values recommended by the Solar Fusion II compilation \cite{RevModPhys.83.195} as well as by the more recent work of S. Daigle \textit{et al.} \cite{PhysRevC.94.025803} (see Table\,\ref{tab:278strength_literature}). We do not quote here a new recommended value, we just note that considering our new value determined with an independent technique, the strength recommended by the Solar Fusion II compilation \cite{RevModPhys.83.195} and especially its somewhat higher uncertainty seems more appropriate than the value of S. Daigle \textit{et al.} \cite{PhysRevC.94.025803} with its very small error bar. The results of those experiments where the $E_p$\,=\,278\,keV resonance is used as a normalization point, do not change by the present result. Therefore, the astrophysical consequences are also unchanged.

The strength of the $E_p$\,=\,1058\,keV resonance, on the other hand, was measured to be significantly higher than the ones determined in the two most recent works (see Table\,\ref{tab:1058strength_literature}). One reason can be that the finite target thickness correction might not have been done in those experiments. In the case of U. Schr\"oder   {\it et al.} \cite{SCHRODER1987240} there is no information about this in the paper. In the case of M. Marta {\it et al.} \cite{PhysRevC.81.055807} it is confirmed that no such a correction has been applied \cite{bemmerer_private}. Based on the information available in \cite{PhysRevC.81.055807} and \cite{marta_thesis}, the correction should be about 7\,\%. This would lead to a resonance strength of $\omega\gamma_{1058}$\,=\,(389\,$\pm$\,22)\,meV. This value still differs from the present one by about two standard deviations \footnote{The subtraction of the non-resonant component in \cite{PhysRevC.81.055807} may also be overestimated. According to  \cite{marta_thesis}, the off-resonance measurement below the resonance was done at 1049\,keV, where the contribution from the low energy tail of the resonance can still be about 5\,\%}. Consequently, as opposed to the $E_p$\,=\,278\,keV resonance, this strength value is rather uncertain and further measurements would be required.

As a summary, the activation technique was successfully used in the present work for the $^{14}$N(p,$\gamma$)$^{15}$O reaction and precise resonance strength values were derived. This technique can also be applied for the measurement of the non-resonant $^{14}$N(p,$\gamma$)$^{15}$O cross section. Such an experiment is in progress using the setup introduced here. The results will be presented in a forthcoming publication. As the activation method provides data which are complementary to the prompt gamma data, the combination of the results can lead to more precise cross section of the $^{14}$N(p,$\gamma$)$^{15}$O astrophysical key reaction.

\begin{acknowledgments}
This work was supported by NKFIH grants K120666 and NN128072, by the \'UNKP-18-4-DE-449 New National Excellence Program of the Human Capacities of Hungary and by the COST Association (ChETEC, CA16117). G.G. Kiss acknowledges support form the J\'anos Bolyai research fellowship of the Hungarian Academy of Sciences. The authors thank I. Rajta, I. Vajda, G. Solt\'esz and Zs. Sz\H ucs for providing excellent beams and working conditions at the Tandetron accelerator. 
\end{acknowledgments}



\begin{thebibliography}{34}
\expandafter\ifx\csname natexlab\endcsname\relax\def\natexlab#1{#1}\fi
\expandafter\ifx\csname bibnamefont\endcsname\relax
  \def\bibnamefont#1{#1}\fi
\expandafter\ifx\csname bibfnamefont\endcsname\relax
  \def\bibfnamefont#1{#1}\fi
\expandafter\ifx\csname citenamefont\endcsname\relax
  \def\citenamefont#1{#1}\fi
\expandafter\ifx\csname url\endcsname\relax
  \def\url#1{\texttt{#1}}\fi
\expandafter\ifx\csname urlprefix\endcsname\relax\def\urlprefix{URL }\fi
\providecommand{\bibinfo}[2]{#2}
\providecommand{\eprint}[2][]{\url{#2}}

\bibitem[{\citenamefont{Wiescher}(2018)}]{Wiescher2018}
\bibinfo{author}{\bibfnamefont{M.}~\bibnamefont{Wiescher}},
  \bibinfo{journal}{Physics in Perspective} \textbf{\bibinfo{volume}{20}},
  \bibinfo{pages}{124} (\bibinfo{year}{2018}), ISSN \bibinfo{issn}{1422-6960},
  \urlprefix\url{https://doi.org/10.1007/s00016-018-0216-0}.

\bibitem[{\citenamefont{Degl'Innocenti
  et~al.}(2004)\citenamefont{Degl'Innocenti, Fiorentini, Ricci, and
  Villante}}]{DEGLINNOCENTI200413}
\bibinfo{author}{\bibfnamefont{S.}~\bibnamefont{Degl'Innocenti}},
  \bibinfo{author}{\bibfnamefont{G.}~\bibnamefont{Fiorentini}},
  \bibinfo{author}{\bibfnamefont{B.}~\bibnamefont{Ricci}}, \bibnamefont{and}
  \bibinfo{author}{\bibfnamefont{F.}~\bibnamefont{Villante}},
  \bibinfo{journal}{Physics Letters B} \textbf{\bibinfo{volume}{590}},
  \bibinfo{pages}{13 } (\bibinfo{year}{2004}), ISSN \bibinfo{issn}{0370-2693},
  \urlprefix\url{http://www.sciencedirect.com/science/article/pii/S0370269304005283}.

\bibitem[{\citenamefont{Fields et~al.}(2018)\citenamefont{Fields, Timmes,
  Farmer, Petermann, Wolf, and Couch}}]{Fields_2018}
\bibinfo{author}{\bibfnamefont{C.~E.} \bibnamefont{Fields}},
  \bibinfo{author}{\bibfnamefont{F.~X.} \bibnamefont{Timmes}},
  \bibinfo{author}{\bibfnamefont{R.}~\bibnamefont{Farmer}},
  \bibinfo{author}{\bibfnamefont{I.}~\bibnamefont{Petermann}},
  \bibinfo{author}{\bibfnamefont{W.~M.} \bibnamefont{Wolf}}, \bibnamefont{and}
  \bibinfo{author}{\bibfnamefont{S.~M.} \bibnamefont{Couch}},
  \bibinfo{journal}{The Astrophysical Journal Supplement Series}
  \textbf{\bibinfo{volume}{234}}, \bibinfo{pages}{19} (\bibinfo{year}{2018}),
  \urlprefix\url{https://doi.org/10.3847%2F1538-4365%2Faaa29b}.

\bibitem[{\citenamefont{Haxton and Serenelli}(2008)}]{Haxton_2008}
\bibinfo{author}{\bibfnamefont{W.~C.} \bibnamefont{Haxton}} \bibnamefont{and}
  \bibinfo{author}{\bibfnamefont{A.~M.} \bibnamefont{Serenelli}},
  \bibinfo{journal}{The Astrophysical Journal} \textbf{\bibinfo{volume}{687}},
  \bibinfo{pages}{678} (\bibinfo{year}{2008}),
  \urlprefix\url{https://doi.org/10.1086%2F591787}.

\bibitem[{\citenamefont{Adelberger et~al.}(2011)\citenamefont{Adelberger,
  Garc\'{\i}a, Robertson, Snover, Balantekin, Heeger, Ramsey-Musolf, Bemmerer,
  Junghans, Bertulani et~al.}}]{RevModPhys.83.195}
\bibinfo{author}{\bibfnamefont{E.~G.} \bibnamefont{Adelberger}},
  \bibinfo{author}{\bibfnamefont{A.}~\bibnamefont{Garc\'{\i}a}},
  \bibinfo{author}{\bibfnamefont{R.~G.~H.} \bibnamefont{Robertson}},
  \bibinfo{author}{\bibfnamefont{K.~A.} \bibnamefont{Snover}},
  \bibinfo{author}{\bibfnamefont{A.~B.} \bibnamefont{Balantekin}},
  \bibinfo{author}{\bibfnamefont{K.}~\bibnamefont{Heeger}},
  \bibinfo{author}{\bibfnamefont{M.~J.} \bibnamefont{Ramsey-Musolf}},
  \bibinfo{author}{\bibfnamefont{D.}~\bibnamefont{Bemmerer}},
  \bibinfo{author}{\bibfnamefont{A.}~\bibnamefont{Junghans}},
  \bibinfo{author}{\bibfnamefont{C.~A.} \bibnamefont{Bertulani}},
  \bibnamefont{et~al.}, \bibinfo{journal}{Rev. Mod. Phys.}
  \textbf{\bibinfo{volume}{83}}, \bibinfo{pages}{195} (\bibinfo{year}{2011}),
  \urlprefix\url{https://link.aps.org/doi/10.1103/RevModPhys.83.195}.

\bibitem[{\citenamefont{Daigle et~al.}(2016)\citenamefont{Daigle, Kelly,
  Champagne, Buckner, Iliadis, and Howard}}]{PhysRevC.94.025803}
\bibinfo{author}{\bibfnamefont{S.}~\bibnamefont{Daigle}},
  \bibinfo{author}{\bibfnamefont{K.~J.} \bibnamefont{Kelly}},
  \bibinfo{author}{\bibfnamefont{A.~E.} \bibnamefont{Champagne}},
  \bibinfo{author}{\bibfnamefont{M.~Q.} \bibnamefont{Buckner}},
  \bibinfo{author}{\bibfnamefont{C.}~\bibnamefont{Iliadis}}, \bibnamefont{and}
  \bibinfo{author}{\bibfnamefont{C.}~\bibnamefont{Howard}},
  \bibinfo{journal}{Phys. Rev. C} \textbf{\bibinfo{volume}{94}},
  \bibinfo{pages}{025803} (\bibinfo{year}{2016}),
  \urlprefix\url{https://link.aps.org/doi/10.1103/PhysRevC.94.025803}.

\bibitem[{\citenamefont{Li et~al.}(2016)\citenamefont{Li, G\"orres, deBoer,
  Imbriani, Best, Kontos, LeBlanc, Uberseder, and
  Wiescher}}]{PhysRevC.93.055806}
\bibinfo{author}{\bibfnamefont{Q.}~\bibnamefont{Li}},
  \bibinfo{author}{\bibfnamefont{J.}~\bibnamefont{G\"orres}},
  \bibinfo{author}{\bibfnamefont{R.~J.} \bibnamefont{deBoer}},
  \bibinfo{author}{\bibfnamefont{G.}~\bibnamefont{Imbriani}},
  \bibinfo{author}{\bibfnamefont{A.}~\bibnamefont{Best}},
  \bibinfo{author}{\bibfnamefont{A.}~\bibnamefont{Kontos}},
  \bibinfo{author}{\bibfnamefont{P.~J.} \bibnamefont{LeBlanc}},
  \bibinfo{author}{\bibfnamefont{E.}~\bibnamefont{Uberseder}},
  \bibnamefont{and} \bibinfo{author}{\bibfnamefont{M.}~\bibnamefont{Wiescher}},
  \bibinfo{journal}{Phys. Rev. C} \textbf{\bibinfo{volume}{93}},
  \bibinfo{pages}{055806} (\bibinfo{year}{2016}),
  \urlprefix\url{https://link.aps.org/doi/10.1103/PhysRevC.93.055806}.

\bibitem[{\citenamefont{Wagner et~al.}(2018)\citenamefont{Wagner, Akhmadaliev,
  Anders, Bemmerer, Caciolli, Gohl, Grieger, Junghans, Marta, Munnik
  et~al.}}]{PhysRevC.97.015801}
\bibinfo{author}{\bibfnamefont{L.}~\bibnamefont{Wagner}},
  \bibinfo{author}{\bibfnamefont{S.}~\bibnamefont{Akhmadaliev}},
  \bibinfo{author}{\bibfnamefont{M.}~\bibnamefont{Anders}},
  \bibinfo{author}{\bibfnamefont{D.}~\bibnamefont{Bemmerer}},
  \bibinfo{author}{\bibfnamefont{A.}~\bibnamefont{Caciolli}},
  \bibinfo{author}{\bibfnamefont{S.}~\bibnamefont{Gohl}},
  \bibinfo{author}{\bibfnamefont{M.}~\bibnamefont{Grieger}},
  \bibinfo{author}{\bibfnamefont{A.}~\bibnamefont{Junghans}},
  \bibinfo{author}{\bibfnamefont{M.}~\bibnamefont{Marta}},
  \bibinfo{author}{\bibfnamefont{F.}~\bibnamefont{Munnik}},
  \bibnamefont{et~al.}, \bibinfo{journal}{Phys. Rev. C}
  \textbf{\bibinfo{volume}{97}}, \bibinfo{pages}{015801}
  (\bibinfo{year}{2018}),
  \urlprefix\url{https://link.aps.org/doi/10.1103/PhysRevC.97.015801}.

\bibitem[{\citenamefont{Gy{\"u}rky et~al.}(2019)\citenamefont{Gy{\"u}rky,
  F{\"u}l{\"o}p, K{\"a}ppeler, Kiss, and Wallner}}]{Gyurky2019}
\bibinfo{author}{\bibfnamefont{G.}~\bibnamefont{Gy{\"u}rky}},
  \bibinfo{author}{\bibfnamefont{Z.}~\bibnamefont{F{\"u}l{\"o}p}},
  \bibinfo{author}{\bibfnamefont{F.}~\bibnamefont{K{\"a}ppeler}},
  \bibinfo{author}{\bibfnamefont{G.~G.} \bibnamefont{Kiss}}, \bibnamefont{and}
  \bibinfo{author}{\bibfnamefont{A.}~\bibnamefont{Wallner}},
  \bibinfo{journal}{The European Physical Journal A}
  \textbf{\bibinfo{volume}{55}}, \bibinfo{pages}{41} (\bibinfo{year}{2019}),
  ISSN \bibinfo{issn}{1434-601X},
  \urlprefix\url{https://doi.org/10.1140/epja/i2019-12708-4}.

\bibitem[{\citenamefont{Woodbury et~al.}(1949)\citenamefont{Woodbury, Hall.,
  and Fowler}}]{woodbury1949}
\bibinfo{author}{\bibfnamefont{E.~J.} \bibnamefont{Woodbury}},
  \bibinfo{author}{\bibfnamefont{R.~N.} \bibnamefont{Hall.}}, \bibnamefont{and}
  \bibinfo{author}{\bibfnamefont{W.~A.} \bibnamefont{Fowler}},
  \bibinfo{journal}{Proceedings of the American Physical Society}
  \textbf{\bibinfo{volume}{75}}, \bibinfo{pages}{1462} (\bibinfo{year}{1949}).

\bibitem[{\citenamefont{Duncan and Perry}(1951)}]{PhysRev.82.809}
\bibinfo{author}{\bibfnamefont{D.~B.} \bibnamefont{Duncan}} \bibnamefont{and}
  \bibinfo{author}{\bibfnamefont{J.~E.} \bibnamefont{Perry}},
  \bibinfo{journal}{Phys. Rev.} \textbf{\bibinfo{volume}{82}},
  \bibinfo{pages}{809} (\bibinfo{year}{1951}),
  \urlprefix\url{https://link.aps.org/doi/10.1103/PhysRev.82.809}.

\bibitem[{\citenamefont{Bashkin et~al.}(1955)\citenamefont{Bashkin, Carlson,
  and Nelson}}]{PhysRev.99.107}
\bibinfo{author}{\bibfnamefont{S.}~\bibnamefont{Bashkin}},
  \bibinfo{author}{\bibfnamefont{R.~R.} \bibnamefont{Carlson}},
  \bibnamefont{and} \bibinfo{author}{\bibfnamefont{E.~B.}
  \bibnamefont{Nelson}}, \bibinfo{journal}{Phys. Rev.}
  \textbf{\bibinfo{volume}{99}}, \bibinfo{pages}{107} (\bibinfo{year}{1955}),
  \urlprefix\url{https://link.aps.org/doi/10.1103/PhysRev.99.107}.

\bibitem[{\citenamefont{Hebbard and Bailey}(1963)}]{HEBBARD1963666}
\bibinfo{author}{\bibfnamefont{D.}~\bibnamefont{Hebbard}} \bibnamefont{and}
  \bibinfo{author}{\bibfnamefont{G.}~\bibnamefont{Bailey}},
  \bibinfo{journal}{Nuclear Physics} \textbf{\bibinfo{volume}{49}},
  \bibinfo{pages}{666 } (\bibinfo{year}{1963}), ISSN \bibinfo{issn}{0029-5582},
  \urlprefix\url{http://www.sciencedirect.com/science/article/pii/0029558263901305}.

\bibitem[{\citenamefont{Becker et~al.}(1982)\citenamefont{Becker, Kieser,
  Rolfs, Trautvetter, and Wiescher}}]{Becker1982}
\bibinfo{author}{\bibfnamefont{H.~W.} \bibnamefont{Becker}},
  \bibinfo{author}{\bibfnamefont{W.~E.} \bibnamefont{Kieser}},
  \bibinfo{author}{\bibfnamefont{C.}~\bibnamefont{Rolfs}},
  \bibinfo{author}{\bibfnamefont{H.~P.} \bibnamefont{Trautvetter}},
  \bibnamefont{and} \bibinfo{author}{\bibfnamefont{M.}~\bibnamefont{Wiescher}},
  \bibinfo{journal}{Zeitschrift f{\"u}r Physik A Atoms and Nuclei}
  \textbf{\bibinfo{volume}{305}}, \bibinfo{pages}{319} (\bibinfo{year}{1982}),
  ISSN \bibinfo{issn}{0939-7922},
  \urlprefix\url{https://doi.org/10.1007/BF01419080}.

\bibitem[{\citenamefont{Runkle et~al.}(2005)\citenamefont{Runkle, Champagne,
  Angulo, Fox, Iliadis, Longland, and Pollanen}}]{PhysRevLett.94.082503}
\bibinfo{author}{\bibfnamefont{R.~C.} \bibnamefont{Runkle}},
  \bibinfo{author}{\bibfnamefont{A.~E.} \bibnamefont{Champagne}},
  \bibinfo{author}{\bibfnamefont{C.}~\bibnamefont{Angulo}},
  \bibinfo{author}{\bibfnamefont{C.}~\bibnamefont{Fox}},
  \bibinfo{author}{\bibfnamefont{C.}~\bibnamefont{Iliadis}},
  \bibinfo{author}{\bibfnamefont{R.}~\bibnamefont{Longland}}, \bibnamefont{and}
  \bibinfo{author}{\bibfnamefont{J.}~\bibnamefont{Pollanen}},
  \bibinfo{journal}{Phys. Rev. Lett.} \textbf{\bibinfo{volume}{94}},
  \bibinfo{pages}{082503} (\bibinfo{year}{2005}),
  \urlprefix\url{https://link.aps.org/doi/10.1103/PhysRevLett.94.082503}.

\bibitem[{\citenamefont{Imbriani et~al.}(2005)\citenamefont{Imbriani,
  Costantini, Formicola, Vomiero, Angulo, Bemmerer, Bonetti, Broggini,
  Confortola, Corvisiero et~al.}}]{Imbriani2005}
\bibinfo{author}{\bibfnamefont{G.}~\bibnamefont{Imbriani}},
  \bibinfo{author}{\bibfnamefont{H.}~\bibnamefont{Costantini}},
  \bibinfo{author}{\bibfnamefont{A.}~\bibnamefont{Formicola}},
  \bibinfo{author}{\bibfnamefont{A.}~\bibnamefont{Vomiero}},
  \bibinfo{author}{\bibfnamefont{C.}~\bibnamefont{Angulo}},
  \bibinfo{author}{\bibfnamefont{D.}~\bibnamefont{Bemmerer}},
  \bibinfo{author}{\bibfnamefont{R.}~\bibnamefont{Bonetti}},
  \bibinfo{author}{\bibfnamefont{C.}~\bibnamefont{Broggini}},
  \bibinfo{author}{\bibfnamefont{F.}~\bibnamefont{Confortola}},
  \bibinfo{author}{\bibfnamefont{P.}~\bibnamefont{Corvisiero}},
  \bibnamefont{et~al.}, \bibinfo{journal}{The European Physical Journal A -
  Hadrons and Nuclei} \textbf{\bibinfo{volume}{25}}, \bibinfo{pages}{455}
  (\bibinfo{year}{2005}), ISSN \bibinfo{issn}{1434-601X},
  \urlprefix\url{https://doi.org/10.1140/epja/i2005-10138-7}.

\bibitem[{\citenamefont{Bemmerer et~al.}(2006)\citenamefont{Bemmerer,
  Confortola, Lemut, Bonetti, Broggini, Corvisiero, Costantini, Cruz,
  Formicola, F{\"u}l{\"o}p et~al.}}]{BEMMERER2006297}
\bibinfo{author}{\bibfnamefont{D.}~\bibnamefont{Bemmerer}},
  \bibinfo{author}{\bibfnamefont{F.}~\bibnamefont{Confortola}},
  \bibinfo{author}{\bibfnamefont{A.}~\bibnamefont{Lemut}},
  \bibinfo{author}{\bibfnamefont{R.}~\bibnamefont{Bonetti}},
  \bibinfo{author}{\bibfnamefont{C.}~\bibnamefont{Broggini}},
  \bibinfo{author}{\bibfnamefont{P.}~\bibnamefont{Corvisiero}},
  \bibinfo{author}{\bibfnamefont{H.}~\bibnamefont{Costantini}},
  \bibinfo{author}{\bibfnamefont{J.}~\bibnamefont{Cruz}},
  \bibinfo{author}{\bibfnamefont{A.}~\bibnamefont{Formicola}},
  \bibinfo{author}{\bibfnamefont{Z.}~\bibnamefont{F{\"u}l{\"o}p}},
  \bibnamefont{et~al.}, \bibinfo{journal}{Nuclear Physics A}
  \textbf{\bibinfo{volume}{779}}, \bibinfo{pages}{297 } (\bibinfo{year}{2006}),
  ISSN \bibinfo{issn}{0375-9474},
  \urlprefix\url{http://www.sciencedirect.com/science/article/pii/S0375947406005902}.

\bibitem[{\citenamefont{Schr{\"o}der et~al.}(1987)\citenamefont{Schr{\"o}der,
  Becker, Bogaert, G{\"o}rres, Rolfs, Trautvetter, Azuma, Campbell, King, and
  Vise}}]{SCHRODER1987240}
\bibinfo{author}{\bibfnamefont{U.}~\bibnamefont{Schr{\"o}der}},
  \bibinfo{author}{\bibfnamefont{H.}~\bibnamefont{Becker}},
  \bibinfo{author}{\bibfnamefont{G.}~\bibnamefont{Bogaert}},
  \bibinfo{author}{\bibfnamefont{J.}~\bibnamefont{G{\"o}rres}},
  \bibinfo{author}{\bibfnamefont{C.}~\bibnamefont{Rolfs}},
  \bibinfo{author}{\bibfnamefont{H.}~\bibnamefont{Trautvetter}},
  \bibinfo{author}{\bibfnamefont{R.}~\bibnamefont{Azuma}},
  \bibinfo{author}{\bibfnamefont{C.}~\bibnamefont{Campbell}},
  \bibinfo{author}{\bibfnamefont{J.}~\bibnamefont{King}}, \bibnamefont{and}
  \bibinfo{author}{\bibfnamefont{J.}~\bibnamefont{Vise}},
  \bibinfo{journal}{Nuclear Physics A} \textbf{\bibinfo{volume}{467}},
  \bibinfo{pages}{240 } (\bibinfo{year}{1987}), ISSN \bibinfo{issn}{0375-9474},
  \urlprefix\url{http://www.sciencedirect.com/science/article/pii/0375947487905288}.

\bibitem[{\citenamefont{Marta et~al.}(2010)\citenamefont{Marta, Trompler,
  Bemmerer, Beyer, Broggini, Caciolli, Erhard, F\"ul\"op, Grosse, Gy\"urky
  et~al.}}]{PhysRevC.81.055807}
\bibinfo{author}{\bibfnamefont{M.}~\bibnamefont{Marta}},
  \bibinfo{author}{\bibfnamefont{E.}~\bibnamefont{Trompler}},
  \bibinfo{author}{\bibfnamefont{D.}~\bibnamefont{Bemmerer}},
  \bibinfo{author}{\bibfnamefont{R.}~\bibnamefont{Beyer}},
  \bibinfo{author}{\bibfnamefont{C.}~\bibnamefont{Broggini}},
  \bibinfo{author}{\bibfnamefont{A.}~\bibnamefont{Caciolli}},
  \bibinfo{author}{\bibfnamefont{M.}~\bibnamefont{Erhard}},
  \bibinfo{author}{\bibfnamefont{Z.}~\bibnamefont{F\"ul\"op}},
  \bibinfo{author}{\bibfnamefont{E.}~\bibnamefont{Grosse}},
  \bibinfo{author}{\bibfnamefont{G.}~\bibnamefont{Gy\"urky}},
  \bibnamefont{et~al.}, \bibinfo{journal}{Phys. Rev. C}
  \textbf{\bibinfo{volume}{81}}, \bibinfo{pages}{055807}
  (\bibinfo{year}{2010}),
  \urlprefix\url{https://link.aps.org/doi/10.1103/PhysRevC.81.055807}.

\bibitem[{\citenamefont{Ajzenberg-Selove}(1991)}]{AJZENBERGSELOVE19911}
\bibinfo{author}{\bibfnamefont{F.}~\bibnamefont{Ajzenberg-Selove}},
  \bibinfo{journal}{Nuclear Physics A} \textbf{\bibinfo{volume}{523}},
  \bibinfo{pages}{1 } (\bibinfo{year}{1991}), ISSN \bibinfo{issn}{0375-9474},
  \urlprefix\url{http://www.sciencedirect.com/science/article/pii/037594749190446D}.

\bibitem[{\citenamefont{{Formicola} et~al.}(2004)\citenamefont{{Formicola},
  {Imbriani}, {Costantini}, {Angulo}, {Bemmerer}, {Bonetti}, {Broggini},
  {Corvisiero}, {Cruz}, {Descouvemont} et~al.}}]{2004PhLB..591...61F}
\bibinfo{author}{\bibfnamefont{A.}~\bibnamefont{{Formicola}}},
  \bibinfo{author}{\bibfnamefont{G.}~\bibnamefont{{Imbriani}}},
  \bibinfo{author}{\bibfnamefont{H.}~\bibnamefont{{Costantini}}},
  \bibinfo{author}{\bibfnamefont{C.}~\bibnamefont{{Angulo}}},
  \bibinfo{author}{\bibfnamefont{D.}~\bibnamefont{{Bemmerer}}},
  \bibinfo{author}{\bibfnamefont{R.}~\bibnamefont{{Bonetti}}},
  \bibinfo{author}{\bibfnamefont{C.}~\bibnamefont{{Broggini}}},
  \bibinfo{author}{\bibfnamefont{P.}~\bibnamefont{{Corvisiero}}},
  \bibinfo{author}{\bibfnamefont{J.}~\bibnamefont{{Cruz}}},
  \bibinfo{author}{\bibfnamefont{P.}~\bibnamefont{{Descouvemont}}},
  \bibnamefont{et~al.}, \bibinfo{journal}{Physics Letters B}
  \textbf{\bibinfo{volume}{591}}, \bibinfo{pages}{61} (\bibinfo{year}{2004}).

\bibitem[{\citenamefont{Oechsner}(1993)}]{OECHSNER1993250}
\bibinfo{author}{\bibfnamefont{H.}~\bibnamefont{Oechsner}},
  \bibinfo{journal}{Applied Surface Science} \textbf{\bibinfo{volume}{70-71}},
  \bibinfo{pages}{250 } (\bibinfo{year}{1993}), ISSN \bibinfo{issn}{0169-4332},
  \urlprefix\url{http://www.sciencedirect.com/science/article/pii/016943329390437G}.

\bibitem[{\citenamefont{Vad et~al.}(2009)\citenamefont{Vad, Cs{\'i}k, and
  Langer}}]{vad2009}
\bibinfo{author}{\bibfnamefont{K.}~\bibnamefont{Vad}},
  \bibinfo{author}{\bibfnamefont{A.}~\bibnamefont{Cs{\'i}k}}, \bibnamefont{and}
  \bibinfo{author}{\bibfnamefont{G.}~\bibnamefont{Langer}},
  \bibinfo{journal}{Spectroscopy Europe} \textbf{\bibinfo{volume}{21}},
  \bibinfo{pages}{13} (\bibinfo{year}{2009}), ISSN \bibinfo{issn}{0966-0941}.

\bibitem[{\citenamefont{Husz{\'a}nk et~al.}(2016)\citenamefont{Husz{\'a}nk,
  Csedreki, Kert{\'e}sz, and T{\"o}r{\"o}k}}]{Huszank2016}
\bibinfo{author}{\bibfnamefont{R.}~\bibnamefont{Husz{\'a}nk}},
  \bibinfo{author}{\bibfnamefont{L.}~\bibnamefont{Csedreki}},
  \bibinfo{author}{\bibfnamefont{Z.}~\bibnamefont{Kert{\'e}sz}},
  \bibnamefont{and}
  \bibinfo{author}{\bibfnamefont{Z.}~\bibnamefont{T{\"o}r{\"o}k}},
  \bibinfo{journal}{Journal of Radioanalytical and Nuclear Chemistry}
  \textbf{\bibinfo{volume}{307}}, \bibinfo{pages}{341} (\bibinfo{year}{2016}),
  ISSN \bibinfo{issn}{1588-2780},
  \urlprefix\url{https://doi.org/10.1007/s10967-015-4102-9}.

\bibitem[{\citenamefont{Mayer}()}]{SIMNRA}
\bibinfo{author}{\bibfnamefont{M.}~\bibnamefont{Mayer}},
  \bibinfo{note}{\,SIMNRA version 6.06},
  \urlprefix\url{http://home.mpcdf.mpg.de/~mam/}.

\bibitem[{\citenamefont{Rajta et~al.}(2018)\citenamefont{Rajta, Vajda,
  Gy{\"u}rky, Csedreki, Kiss, Biri, van Oosterhout, Podaru, and
  Mous}}]{RAJTA2018125}
\bibinfo{author}{\bibfnamefont{I.}~\bibnamefont{Rajta}},
  \bibinfo{author}{\bibfnamefont{I.}~\bibnamefont{Vajda}},
  \bibinfo{author}{\bibfnamefont{G.}~\bibnamefont{Gy{\"u}rky}},
  \bibinfo{author}{\bibfnamefont{L.}~\bibnamefont{Csedreki}},
  \bibinfo{author}{\bibfnamefont{Z.}~\bibnamefont{Kiss}},
  \bibinfo{author}{\bibfnamefont{S.}~\bibnamefont{Biri}},
  \bibinfo{author}{\bibfnamefont{H.}~\bibnamefont{van Oosterhout}},
  \bibinfo{author}{\bibfnamefont{N.}~\bibnamefont{Podaru}}, \bibnamefont{and}
  \bibinfo{author}{\bibfnamefont{D.}~\bibnamefont{Mous}},
  \bibinfo{journal}{Nuclear Instruments and Methods in Physics Research Section
  A: Accelerators, Spectrometers, Detectors and Associated Equipment}
  \textbf{\bibinfo{volume}{880}}, \bibinfo{pages}{125 } (\bibinfo{year}{2018}),
  ISSN \bibinfo{issn}{0168-9002},
  \urlprefix\url{http://www.sciencedirect.com/science/article/pii/S0168900217311622}.

\bibitem[{\citenamefont{Gy{\"u}rky et~al.}(2017)\citenamefont{Gy{\"u}rky,
  Ornelas, F{\"u}l{\"o}p, Hal\'asz, Kiss, Sz{\"u}cs, Husz\'ank, Horny\'ak,
  Rajta, and Vajda}}]{PhysRevC.95.035805}
\bibinfo{author}{\bibfnamefont{G.}~\bibnamefont{Gy{\"u}rky}},
  \bibinfo{author}{\bibfnamefont{A.}~\bibnamefont{Ornelas}},
  \bibinfo{author}{\bibfnamefont{Z.}~\bibnamefont{F{\"u}l{\"o}p}},
  \bibinfo{author}{\bibfnamefont{Z.}~\bibnamefont{Hal\'asz}},
  \bibinfo{author}{\bibfnamefont{G.~G.} \bibnamefont{Kiss}},
  \bibinfo{author}{\bibfnamefont{T.}~\bibnamefont{Sz{\"u}cs}},
  \bibinfo{author}{\bibfnamefont{R.}~\bibnamefont{Husz\'ank}},
  \bibinfo{author}{\bibfnamefont{I.}~\bibnamefont{Horny\'ak}},
  \bibinfo{author}{\bibfnamefont{I.}~\bibnamefont{Rajta}}, \bibnamefont{and}
  \bibinfo{author}{\bibfnamefont{I.}~\bibnamefont{Vajda}},
  \bibinfo{journal}{Phys. Rev. C} \textbf{\bibinfo{volume}{95}},
  \bibinfo{pages}{035805} (\bibinfo{year}{2017}),
  \urlprefix\url{https://link.aps.org/doi/10.1103/PhysRevC.95.035805}.

\bibitem[{\citenamefont{Fowler et~al.}(1948)\citenamefont{Fowler, Lauritsen,
  and Lauritsen}}]{RevModPhys.20.236}
\bibinfo{author}{\bibfnamefont{W.~A.} \bibnamefont{Fowler}},
  \bibinfo{author}{\bibfnamefont{C.~C.} \bibnamefont{Lauritsen}},
  \bibnamefont{and}
  \bibinfo{author}{\bibfnamefont{T.}~\bibnamefont{Lauritsen}},
  \bibinfo{journal}{Rev. Mod. Phys.} \textbf{\bibinfo{volume}{20}},
  \bibinfo{pages}{236} (\bibinfo{year}{1948}),
  \urlprefix\url{https://link.aps.org/doi/10.1103/RevModPhys.20.236}.

\bibitem[{\citenamefont{Borowski et~al.}(2009)\citenamefont{Borowski, Lieb,
  Uhrmacher, and Bolse}}]{doi:10.1063/1.3087064}
\bibinfo{author}{\bibfnamefont{M.}~\bibnamefont{Borowski}},
  \bibinfo{author}{\bibfnamefont{K.~P.} \bibnamefont{Lieb}},
  \bibinfo{author}{\bibfnamefont{M.}~\bibnamefont{Uhrmacher}},
  \bibnamefont{and} \bibinfo{author}{\bibfnamefont{W.}~\bibnamefont{Bolse}},
  \bibinfo{journal}{AIP Conference Proceedings}
  \textbf{\bibinfo{volume}{1090}}, \bibinfo{pages}{450} (\bibinfo{year}{2009}),
  \urlprefix\url{https://aip.scitation.org/doi/abs/10.1063/1.3087064}.

\bibitem[{\citenamefont{Ziegler}()}]{SRIM}
\bibinfo{author}{\bibfnamefont{J.}~\bibnamefont{Ziegler}},
  \bibinfo{note}{\,SRIM-2013 software code}, \urlprefix\url{http://srim.org/}.

\bibitem[{\citenamefont{Sakamoto et~al.}(2000)\citenamefont{Sakamoto, Ogawa,
  and Tsuchida}}]{SAKAMOTO2000250}
\bibinfo{author}{\bibfnamefont{N.}~\bibnamefont{Sakamoto}},
  \bibinfo{author}{\bibfnamefont{H.}~\bibnamefont{Ogawa}}, \bibnamefont{and}
  \bibinfo{author}{\bibfnamefont{H.}~\bibnamefont{Tsuchida}},
  \bibinfo{journal}{Nuclear Instruments and Methods in Physics Research Section
  B: Beam Interactions with Materials and Atoms}
  \textbf{\bibinfo{volume}{164-165}}, \bibinfo{pages}{250 }
  (\bibinfo{year}{2000}), ISSN \bibinfo{issn}{0168-583X},
  \urlprefix\url{http://www.sciencedirect.com/science/article/pii/S0168583X99010885}.

\bibitem[{\citenamefont{Reiter et~al.}(1987)\citenamefont{Reiter, Baumgart,
  Kniest, Pfaff, and Clausnitzer}}]{REITER1987287}
\bibinfo{author}{\bibfnamefont{G.}~\bibnamefont{Reiter}},
  \bibinfo{author}{\bibfnamefont{H.}~\bibnamefont{Baumgart}},
  \bibinfo{author}{\bibfnamefont{N.}~\bibnamefont{Kniest}},
  \bibinfo{author}{\bibfnamefont{E.}~\bibnamefont{Pfaff}}, \bibnamefont{and}
  \bibinfo{author}{\bibfnamefont{G.}~\bibnamefont{Clausnitzer}},
  \bibinfo{journal}{Nuclear Instruments and Methods in Physics Research Section
  B: Beam Interactions with Materials and Atoms} \textbf{\bibinfo{volume}{27}},
  \bibinfo{pages}{287 } (\bibinfo{year}{1987}), ISSN \bibinfo{issn}{0168-583X},
  \urlprefix\url{http://www.sciencedirect.com/science/article/pii/0168583X87905672}.

\bibitem[{\citenamefont{Bemmerer}()}]{bemmerer_private}
\bibinfo{author}{\bibfnamefont{D.}~\bibnamefont{Bemmerer}},
  \bibinfo{note}{private communication}.

\bibitem[{\citenamefont{Marta}(2011)}]{marta_thesis}
\bibinfo{author}{\bibfnamefont{M.}~\bibnamefont{Marta}}, Ph.D. thesis,
  \bibinfo{school}{TU Dresden, Germany} (\bibinfo{year}{2011}).

\end{thebibliography}
\end{document}